\documentstyle[psfig,twocolumn,prl,aps,amssymb]{revtex}
\begin{document}
\wideabs{
\draft

\title{Spontaneous Magnetization of Composite Fermions}
\author{K. Park and J. K. Jain}
\address{Department of Physics, 104 Davey Laboratory,
The Pennsylvania State University,
Pennsylvania 16802}
\date{\today}

\maketitle

\begin{abstract}

It is argued that the composite fermion liquid is a promising candidate for an
observation of the elusive, interaction driven magnetization 
first proposed by Bloch seven decades ago.  In analogy to what is 
theoretically believed to be the case for the idealized electron gas 
in zero magnetic field, this spontaneously broken symmetry 
phase is predicted to occur prior to a transition into the Wigner crystal. 

\end{abstract}

}

The system of electrons in a uniform, positively charged background 
is a  widely used and studied model in condensed matter physics. 
It was suggested by Bloch in 1929 \cite {Bloch}
that the Fermi sea of electrons 
is susceptible to a spontaneous polarization of the
electron spin at low densities, when the interaction 
becomes strong relative to the kinetic energy.  The physical origin is
obvious:  At sufficiently low densities, electrons gain more in exchange 
energy by aligning their spins than they lose in kinetic energy.
Another competing  phase at low-densities is the Wigner crystal (WC), 
a lattice of electrons.  A phase diagram for interacting electrons as a function 
of density has been the subject of much theoretical study and controversy.
The two dimensional (2D) electron system, for example that at the 
interface of two semiconductors, 
is of interest since it is a rather faithful realization of the
idealized jellium model with a rigid, uniform positive background.  
The usual perturbative approaches, e.g.
the Hartree-Fock or random-phase approximation, are not useful at 
low densities; sophisticated quantum Monte Carlo calculations 
indicate that a transition into the WC state occurs at
$r_s\approx 37$ \cite {Ceperley}, 
preceded by a ferromagnetic phase \cite {Senatore}. 
The ferromagnetic Bloch phase has not yet been observed in 
2D electron systems, though. (For recent theoretical and experimental work 
on three dimensional systems, see Refs.~\cite{Ortiz}.)

In recent years, there has been intense 
interest in various states of a new species of
fermions, called composite fermions \cite{book1,book2,Jain}, 
which are formed when electrons
confined to two dimensions are exposed to a strong magnetic field.  Although
of a collective origin, composite fermions behave like ordinary fermions 
in many respects.  In particular, their Fermi sea, \cite {HLR}, 
cyclotron orbits, Shubnikov-de Haas oscillations, and  quantized Landau 
levels (LLs) have been observed experimentally \cite{book1,book2}.      
It is also known that a transition from the composite fermion (CF)
liquid to the WC takes place at
small Landau level fillings \cite{Lam}.  Motivated by the above discussion,
we have searched for a spontaneous magnetic ordering in the CF 
liquid in the vicinity of the WC phase transition.

Composite fermions are bound states of electrons and an even number 
of magnetic flux quanta \cite {Comment}, with a flux quantum defined as 
$\phi_0=hc/e$, and are formed in the lowest LL because
electrons best screen the repulsive Coulomb interaction by capturing $2p$ flux
quanta and transforming into composite fermions. 
The most fundamental property of composite fermions  
is that they experience a much reduced effective 
magnetic field compared to electrons, given by $B^*=B-2p\rho\phi_0$, where
$B$ is the external magnetic field and $\rho$ is the density.
In effect, the captured flux, given by the last term in the expression for $B^*$,
becomes invisible to composite fermions.  
The effective filling factor of composite fermions, $\nu^*=\rho\phi_0/|B^*|$, is
related to the electron filling factor, $\nu=\rho\phi_0/B$, by 
$\nu=\nu^*/(2p\nu^*\pm 1)$.  Composite fermions exhibit 
integer quantum Hall effect (IQHE)
 when the effective field is such that an integer number of CF-LLs
are occupied, with $n$ filled CF-LLs corresponding to fractional QHE (FQHE) of
electrons at $\nu=n/(2pn\pm 1)$, and form a composite Fermi sea at electron
filling factors $\nu=1/2p$, where the effective field vanishes.  
The CF state is in general not fully spin polarized and has  
$n$$\uparrow$ ($n$$\downarrow$) 
spin-up (spin-down) CF-LLs occupied ($n=n$$\uparrow+n$$\downarrow$).
The corresponding wave function is given by \cite {Jain} 
\begin{equation}
\Phi^J_{\frac{n}{2pn+1}}={\cal P} \prod_{j<k=1}^N(z_j-z_k)^{2p}
A[\Phi_{n\uparrow} \Phi_{n\downarrow}u_1 ...
u_{N\uparrow}d_{N\uparrow+1}...d_N]
\end{equation}
where $\Phi_{n\uparrow}$ ($\Phi_{n\downarrow}$) is the Slater determinant
wave function of $N$$\uparrow$ spin-up ($N$$\downarrow=N-N$$\uparrow$ spin-down) 
electrons occupying $n$$\uparrow$
($n$$\downarrow$) LLs, $u$ and $d$ are the up and down spinors, $A$ is the
antisymmetrization operator, $z_j=x_j+iy_j$ is the position of the $j$th
particle, and the operator ${\cal P}$ projects the wave function into
the lowest electronic LL.  $\Phi^J$ obviously 
does not contain any adjustable parameters.
We will denote the composite fermions carrying $2p$ flux quanta by $^{2p}$CFs and
the state with $n$ filled $^{2p}$CF-LLs by $^{2p}$CF$_n$, or by
$^{2p}$CF$_{n\uparrow,n\downarrow}$ when the polarization is of interest.
The $^{2p}$CFs are relevant in the filling factor range
$\frac{1}{2p-1}>\nu\geq \frac{1}{2p+1}$.

It has been known that $\Phi^J$ provides  
an accurate quantum mechanical 
description of the actual state, whether
fully or non-fully polarized \cite{Dev,Wu,JK},
obtaining energies correctly to within 0.1\% for all cases 
where exact results are known from numerical diagonalization.  
We will be concerned here with an effect that is caused by the
residual interaction between composite fermions, which, by definition, 
is whatever remains after most of the Coulomb
interaction is used up in giving the composite fermion its mass.
$\Phi^J$ also captures subtle effects originating from the residual
inter-CF interaction.  For example, the energy splittings between
various CF states which would be degenerate for non-interacting composite
fermions  are predicted extremely accurately.
In particular, the dispersion of the CF exciton 
(for non-interacting composite fermions, the exciton 
will have a constant energy) is obtained accurately, to the extent 
that it explicitly shows a CDW instability 
toward Wigner crystallization at small $\nu$ \cite {JK}.  
Of particular interest here is the reliability of the CF theory in predicting
the spin-polarizations of various QHE states, which have been studied in detail
for $^2$CFs, relevant in the filling factor range $2/3>\nu>1/3$.
The theoretical phase diagram of the spin polarization
of $^2$CF$_n$ as a function of the Zeeman energy
computed with the help of $\Phi^J$ \cite{Park} is in reasonably good 
quantitative agreement with the experimentally determined 
phase diagrams \cite {Du,Kukushkin}.
For $^2$CFs, it turns out that the model of {\em independent}
composite fermions is successful in predicting various qualitative features, 
namely the possible spin polarizations as well as the
energy ordering of the differently polarized states; in particular, the 
ground state in the absence of the Zeeman energy is 
the least polarized state, as expected for weakly interacting fermions.

\begin{figure}
\centerline{\psfig{figure=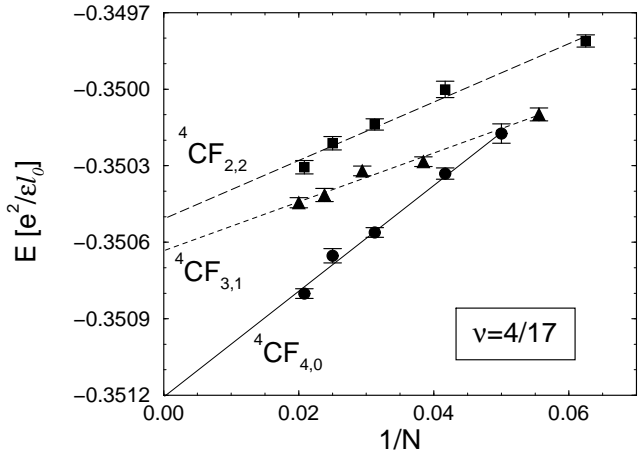,width=4.0in,angle=-90}}
\caption{Thermodynamic extrapolations  
for the energies of $^4$CF$_{2,2}$, $^4$CF$_{3,1}$, and $^4$CF$_{4,0}$, the 
variously spin-polarized states
of $^4$CFs at 4/17.  The energies are quoted in units of $e^2/\epsilon l$, where
$l=\sqrt{\hbar c/eB}$ is the magnetic length and $\epsilon$ is the dielectric
constant of the background material.  The lines show the best straight line fits.
\label{fig1}}
\end{figure}

\begin{figure}
\centerline{\psfig{figure=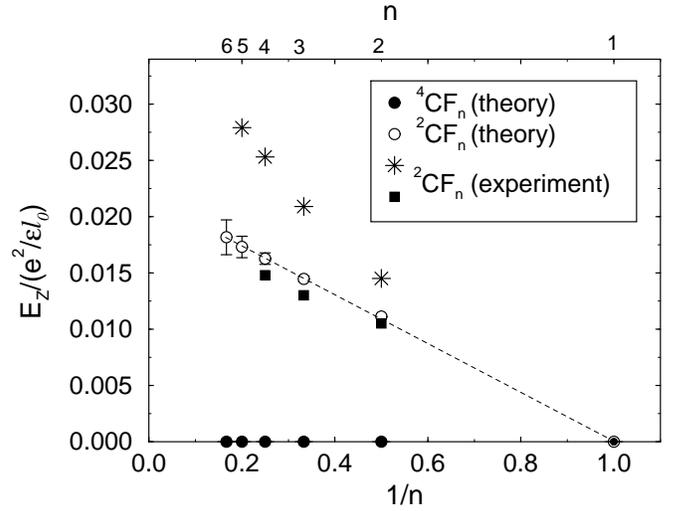,width=4.0in,angle=-90}}
\caption{The critical Zeeman energy above which the CF liquid is fully polarized
for $^4$CFs (filled circles).  For comparison, the theoretical 
critical Zeeman energy is also shown for
$^2$CFs, taken from Ref.~\protect \cite {Park} (open circles), 
along with the experimental results 
from Ref.~\protect \cite {Du} (stars) and Ref.~\protect\cite{Kukushkin} (squares).
\label{fig2}}
\end{figure}

As stated above, we search for Bloch's magnetization prior to Wigner
crystallization.  Of interest here is the {\em intrinsic}
magnetic ordering caused by the
interaction, and not the trivial magnetization due to
the Zeeman coupling of the electron magnetic moment to the
external magnetic field.  The Zeeman coupling will therefore be set
to zero in what follows; the relevance of the results to  
experiment will be discussed below.
The relevant parameter here is the filling factor, $\nu$, with small $\nu$
analogous to low density.
Since there is experimental evidence for the WC phase 
on both sides of $^4$CF$_1$ (i.e., $\nu=1/5$) \cite{WC}, 
we focus on $^4$CFs and evaluate the energies of
$^4$CF$_{n\uparrow,n\downarrow}$ by quantum Monte Carlo technique, with
the lowest LL projection handled as discussed in Ref.~\cite {JK}. 
The kinetic energy is quenched in the lowest Landau level, 
and the (non-relativistic) Hamiltonian is simply:
\begin{equation}
H=\frac{1}{2}\sum_{j\neq k}\frac{e^2}{\epsilon|{\bf r}_j-{\bf r}_k|} + V_{e-b}
\end{equation} 
where ${\bf r}_{j}$ is the position of the $j$th particle, $V_{e-b}$
is the interaction of electrons with the uniform positively charged
background, and $\epsilon$ is the background dielectric constant.  As we are
interested in thermodynamic phases,  all energies are obtained by a careful
extrapolation to the thermodynamic limit, 
$N^{-1}\rightarrow 0$, as shown in Fig.~(\ref{fig1}) for $^4$CF$_{4,0}$,
$^4$CF$_{3,1}$, and  $^4$CF$_{2,2}$ .  A consideration of
large systems as well as 
extrapolation to $N^{-1}\rightarrow 0$ is crucial, since the
ordering of states often changes as a function of $N$, as also noted earlier \cite
{Xie}.  The energy differences are as small as 0.03\% of the ground state 
energy, and it requires up to $10^7$ Monte Carlo steps to obtain each energy 
with sufficient accuracy.  The principal result of this work is that 
the energy ordering of the $^4$CF states of different 
polarizations is opposite to that of the $^2$CFs states, 
with the fully polarized state now being the ground state.
The model of independent composite fermions thus dramatically fails  
for $^4$CFs at small Zeeman energies, indicating that the 
inter-CF interaction is  
sufficiently strong to cause a spontaneous magnetization of the $^4$CF liquid.  
The phase diagram of the spin polarization of $^4$CF$_n$ is contrasted with that
of $^2$CF$_n$ in Fig.~(\ref{fig2}).

\begin{figure}
\centerline{\psfig{figure=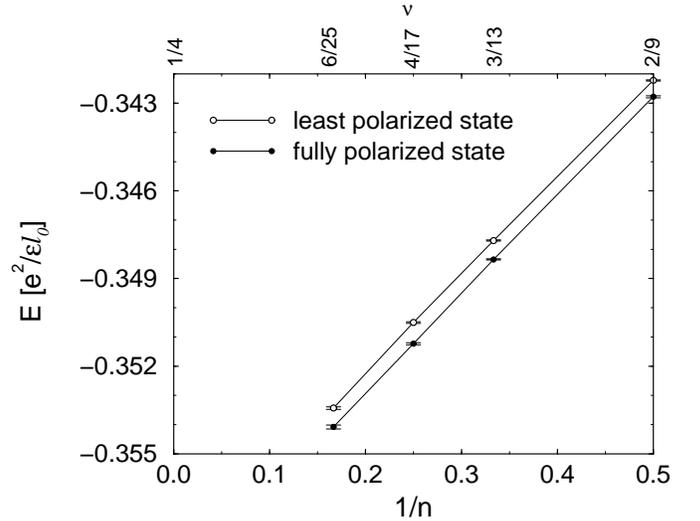,width=4.0in,angle=-90}}
\caption{The energies of the fully and the least polarized 
$^4$CF$_n$ states at $\nu=n/(4n+1)$.  The lines are a guide to the eye.
\label{fig3}}
\end{figure}

The spin-polarization of the $^4$CF sea at $\nu=1/4$, 
$^4$CF$_\infty$, is also of interest.  The results in 
Fig.~(\ref{fig3}) indicate that the energy 
difference between the fully and the least
polarized states is to a good degree 
independent of $n$, suggesting that 
$^4$CF$_\infty$ is at least partially spin polarized.  Further, since  
we find that the fully polarized state remains the ground state for up to
$n=6$ ($\nu=6/25$), we suspect that $^4$CF$_\infty$ is close to fully polarized;
it is, however, not possible to estimate the actual value of magnetization 
from our calculations.

There exists evidence \cite {Dumass} that the effective mass of
composite fermion increases for $^2$CFs as one approaches the half-filled
LL along the sequence $n/(2n+1)$.  One might expect that this would 
make the kinetic energy of composite fermions
less important, causing an intrinsic magnetic ordering  
for $^2$CF$_\infty$ at $\nu=\frac{1}{2}$ as well.  A calculation of the  
energies of $^2$CF$_{n\uparrow,n\downarrow}$ 
shows that the unpolarized state remains the ground state 
at least for up to $n=n$$\uparrow+n$$\downarrow=8$ ($\nu=8/17$) \cite {Comment2}.  
This strongly suggests that $^2$CF$_\infty$ is unpolarized 
in the absence of Zeeman energy, although a {\em weak} intrinsic 
ferromagnetism at $\nu=\frac{1}{2}$ can not be ruled out.

\begin{figure}
\centerline{\psfig{figure=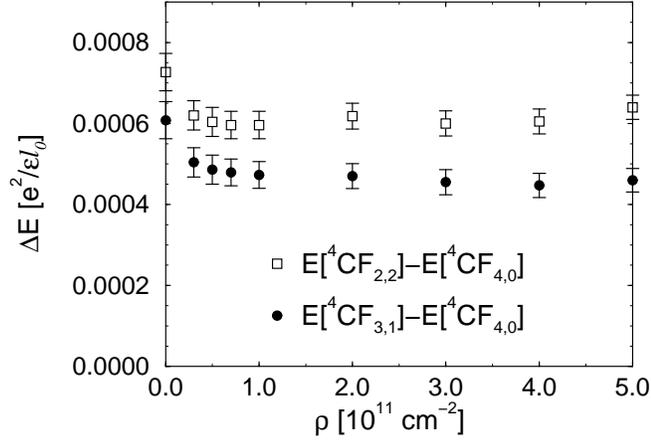,width=4.0in,angle=-90}}
\caption{The energy differences between $^4$CF$_{2,2}$ and $^4$CF$_{4,0}$, and
between $^4$CF$_{3,1}$ and $^4$CF$_{4,0}$ as a function of density for a
heterojunction sample, with the effective 
interaction evaluated in the self-consistent local
density approximation.  In the limit of zero density, the thickness vanishes and 
the effective interaction becomes $e^2/\epsilon r$.
\label{fig4}}
\end{figure}

The above calculations assume a strictly 2D electronic wave function, 
whereas in the experimental systems the electron wave function has a 
non-zero transverse extension.  Due to the finite thickness, the effective 
interaction between electrons becomes softer at short distances, with a 
logarithmic rather than $1/r$ divergence as $r\rightarrow 0$.  The actual form of
the effective interaction depends on the electron density as well as the 
form of the confinement potential, which can therefore serve as useful knobs for 
studying how the phase diagram evolves with varying interaction.
We have determined the form of the transverse wave function in the 
self-consistent local density approximation, by iteratively solving a 
combination of the one-dimensional Poisson and Schr\"odinger equations 
as a function of density \cite {Stern}, obtained therefrom the effective
interaction, and then recomputed the energies of various candidate states.  
While the energy differences are slightly altered 
(lowered by approximately 15\%), the qualitative conclusions remain unchanged, 
as seen in Fig.~(\ref{fig4}).

LL mixing has been neglected in the above, which is a reasonable 
approximation at sufficiently large magnetic fields.  LL mixing is more 
significant in hole type samples, due to the larger mass and 
smaller cyclotron energy of 
holes, and shifts transition into the WC state 
from $\nu\approx \frac{1}{5}$ to $\nu\approx \frac{1}{3}$ in typical 
samples \cite{Shayegan}.  It would be
interesting to investigate if the $^2$CF states are fully polarized for these
hole-type samples, say at $\nu=\frac{2}{5}$.  A proper theoretical
treatment of LL mixing is outside the scope of the present work.

The predictions of this work ought to be experimentally verifiable.
The transitions between QHE states of different polarizations have been 
seen in transport experiments \cite {Du}, and the polarization itself has 
been measured in optical luminescence studies 
\cite{Kukushkin} and also by NMR \cite {Barrett}. 
Since the magnetization we are predicting is to be distinguished from 
that caused by the Zeeman coupling,  
we hope our work will motivate polarization measurements under hydrostatic 
pressure, which can be used to tune the $g$ factor through zero 
\cite{Pressure}.  Our results would imply an absence of any transition at 
finite Zeeman energies at $n/(4n+1)$, and a finite jump in the 
degree of polarization when the $g$ factor changes sign. 
In fact, there already may exist a preliminary evidence for the  
effect predicted here.  In Fig.~3c of Kukushkin, von Klitzing, and Eberl 
\cite {Kukushkin}, the polarization of $^2$CF$_{\infty}$ 
seems to  vanish in the limit $B\rightarrow 0$  (which is also the
limit of vanishing Zeeman energy), but, at least a naive extrapolation of 
the polarization of the $^4$CF sea appears to approach a finite value, 
indicating a non-zero intrinsic magnetization at $\nu=\frac{1}{4}$.

Partially spin polarized states have been observed \cite {Yeh} for CFs near 
$\nu=3/4$, which are conceptually similar, but not related by any exact symmetry
to the $^4$CF states 
considered here.  These are technically more complicated, involving
attachment of four flux quanta in two installments of two 
flux quanta each, one regular and one {\em reverse} \cite {Wu},
separated by a particle hole transformation.  Specifically, the state at
$\frac{3n+2}{4n+3}$ is obtained by 
reverse attachment of two flux quanta ($\nu\rightarrow \frac{\nu}{2\nu-1}$) to
the state at $\frac{3n+2}{2n+1}$, which, in turn, is related to  
$\frac{n}{2n+1}$ by particle hole symmetry ($\nu\rightarrow 2-\nu$). 
Unfortunately, we cannot use the present method to determine the theoretical 
phase diagram of these states.

As mentioned above, the energy differences between different possible states are
extremely small, and the results rely heavily on the accuracy of the 
wave functions $\Phi^J$.  It would be useful in the future to test the robustness
of the above predictions by introducing some variational degree of 
freedom in the wave functions.  In particular, the effect of LL mixing can 
be investigated in a fixed phase diffusion Monte Carlo approach \cite {DMG}.

In summary, we predict that the composite fermion liquid 
exhibits a broken symmetry magnetic phase prior 
to a transition into the Wigner crystal.  This is an example in which the residual
inter-CF interaction qualitatively changes the nature of the state by causing a
phase transition.  This work was supported in part by the
National Science Foundation under grant no. DMR-9615005,
and by the National Center for Supercomputing Applications at the University
of Illinois (Origin 2000).

\begin{thebibliography}{99}
\bibitem{Comment2} For 8/17 
we have only considered $^2$CF$_{4,4}$ and $^2$CF$_{5,3}$.
\end{thebibliography}
\end{document}